%% file: main.tex
\documentclass[runningheads]{llncs}
\usepackage{amsmath,amssymb,amsfonts}
\usepackage{algorithmic}
\usepackage{bm}
\usepackage{booktabs}
\usepackage{cite}
\usepackage{color}
\usepackage{courier}
\usepackage{enumitem}
\usepackage{footnote}
\usepackage{graphicx}
\usepackage{multirow}
\usepackage{subcaption}
\usepackage{soul}
\usepackage{textcomp}
\usepackage{url}
\usepackage{xcolor}
\usepackage{xspace}
\usepackage{hyperref}

\newcommand{\paratitle}[1]{\vspace{1.5ex}\noindent{\small\emph{\textbf{#1}}}}
\newcommand{\ie}{\emph{i.e.,}\xspace}

\newcommand{\eg}{\emph{e.g.,}\xspace}

\newcommand{\ignore}[1]{}

\begin{document}
\title{Leveraging Search History for Improving Person-Job Fit}
\author{Yupeng Hou\inst{1}\thanks{Work done during internship at BOSS Zhipin NLP Center.} \and Xingyu Pan\inst{2} \and Wayne Xin Zhao\inst{1,4}\thanks{Corresponding author.} \and Shuqing Bian\inst{2} \and \\ Yang Song\inst{3} \and Tao Zhang\inst{3} \and Ji-Rong Wen\inst{1,2,4}}

\institute{
Gaoling School of Artificial Intelligence, Renmin University of China, China \email{\{houyupeng,jrwen\}@ruc.edu.cn, batmanfly@gmail.com} \and
School of Information, Renmin University of China, China
\email{xy\_pan@foxmail.com, bianshuqing@ruc.edu.cn} \and
BOSS Zhipin, Beijing, China
\email{\{songyang, kylen.zhang\}@kanzhun.com} \and
Beijing Key Laboratory of Big Data Management and Analysis Methods, China
}
\authorrunning{Y. Hou et al.}
\maketitle              %
\begin{abstract}
    As the core technique of online recruitment platforms, \emph{person-job fit} can improve hiring efficiency by accurately matching job positions with qualified candidates.
    However, existing studies mainly focus on the \emph{recommendation} scenario, while neglecting another important channel  for linking positions with job seekers, \ie \emph{search}. 
    Intuitively, search history contains rich user behavior in job seeking,  reflecting important evidence for job intention of users. 
    
    In this paper,  we present a novel \textbf{S}earch \textbf{H}istory enhanced \textbf{P}erson-\textbf{J}ob \textbf{F}it model, named as \textbf{SHPJF}.
    To utilize both text content from jobs/resumes and search histories from users, we propose two components with different purposes. For \emph{text matching component},   we design a BERT-based text encoder for capturing the semantic interaction between resumes and job descriptions.  For \emph{intention modeling component}, we
    design two kinds of intention modeling approaches based on 
    the Transformer architecture, either based on the click sequence or query text sequence. 
    To  capture underlying job intentions, we further propose an intention clustering technique to identify and summarize the major  intentions from search logs. Extensive experiments on a large real-world recruitment dataset  
    have demonstrated the effectiveness of our approach. 
\keywords{Person-job fit  \and Self-attention \and User intention.}
\end{abstract}

\input{sec-intro.tex}
\input{sec-rel.tex}
\input{sec-def.tex}
\input{sec-model.tex}

\input{sec-exp.tex}

\section{Conclusion}
In this paper, we presented the first study that leveraged search history data for improving the task of person-job fit.  
Our approach developed a basic component based on BERT for capturing the semantic interaction between resumes and job descriptions. As the major technical contribution, we designed a novel intention modeling component that was able to learn and identify the intention of job seekers. It modeled two kinds of sequences, either the click sequence or query text sequence. Furthermore, an intention clustering technique is proposed to accurately capture underlying job intentions.
Extensive experiments on a large recruitment data  
have shown the demonstrated the effectiveness of our approach. 

Besides search history, there are other kinds of side information for person-job fit.
As future work, we will consider developing a more general approach to leverage various kinds of side information in a unified way.

\section*{Acknowledgements}
This work was partially supported by the National Natural Science Foundation of China under Grant No. 61872369 and 61832017,
Beijing Outstanding Young Scientist Program under Grant No. BJJWZYJH012019100020098.
Xin Zhao is the corresponding author.

\bibliographystyle{splncs04.bst}
\bibliography{main}

\end{document}

%% file: sec-intro.tex
\section{Introduction}
With the rapid development of Web techniques,  online recruitment has become prevalent to match qualified candidates with suitable job positions or vacancies through the online service.
As the core technique of online recruitment, it is key to develop an effective algorithm for the task of \emph{person-job fit}~\cite{zhu2018person}, given the huge increase in both job seekers and job positions.

On online recruitment platforms, businesses  publish job descriptions introducing the requirement for the positions, while job seekers upload their resumes stating their skills and the expectation about jobs. 
A job seeker can browse the job listings, send applications to interested positions and further schedule the interviews. Since both job descriptions and candidate resumes are presented in natural language,  person-job fit is typically casted as a text-matching task~\cite{zhu2018person}, and a number of studies~\cite{qin2018enhancing,bian2019domain,bian2020learning} aim to learn effective text representation models for capturing the semantic compatibility between job and resume contents. 

Existing studies mainly focus on the \emph{recommendation} scenario, where the system provides recommended job positions and a job seeker provides implicit or explicit feedback to these recommendations through  behaviors such as clicking or making an application. However, another important channel of \emph{search} has seldom been considered in previous studies: a user can also actively issue queries and click interested jobs like in general-purpose search engine.
Although on online recruitment platforms \emph{recommendation} takes a higher volume of user behaviors, 
\emph{search} also serves an important complementary function. 
According to the statistics on \textcolor{black}{the collected dataset},
search takes a considerable proportion of 19\% volume of user behaviors against the major proportion of 81\% by recommendation. Intuitively, search history contains rich user behavior in job seeking,  reflecting important evidence for job intention of users. In particular, it will be more important to consider when resume text is  not well-written (noisy, short or informal), or when the intention of the job seeker is unclear (\eg low-skilled job seekers tend to consider a diverse range of job categories).

Considering the importance of search history data, we aim to leverage these valuable user data for improving the task of person-job fit. However, there are two major challenges to effectively utilize search history data. First, search history is usually presented in the form of short queries together with clicked (or applied) jobs, and it is unclear how to capture actual job needs from search logs.
Second, the semantics reflected in search history can be redundant or diverse, and it is difficult to find out the  underlying major intentions for job seekers. 

To this end, in this paper, we propose a novel \textbf{S}earch \textbf{H}istory enhanced \textbf{P}erson-\textbf{J}ob \textbf{F}it model, named as \textbf{SHPJF}. It jointly utilizes text content from job descriptions/resumes and search histories from users. On one hand, we design a BERT-based text encoder for capturing the semantic interaction between resumes and job descriptions,
called \emph{text matching component}. On the other hand, we leverage the search history to model the intention of a candidate, called \emph{intention modeling component}.
In intention modeling component, we design two kinds of intention modeling approaches based on 
the Transformer architecture. These two approaches consider modeling the click sequence and query text sequence, respectively. 
To  capture underlying job intentions, we further propose an intention clustering technique to identify and summarize the major  intentions from search logs. 
Finally, we integrate the above two components together.

To the best of our knowledge, it is the first time that search history data has been leveraged to improve the task of person-job fit. To evaluate our approach, we construct a large real-world dataset from a popular \textcolor{black}{online} recruitment platform. Extensive experiments have shown that our approach is more effective than a number of competitive baseline methods. 

%% file: sec-rel.tex
\section{Related Work}

Person-Job Fit (PJF) has been extensively studied in the literature
as an important task in recruitment data mining~\cite{kenthapadi2017personalized, shalaby2017help}.
The early research efforts of person-job fit can be dated back to Malinowski \textit{et al.}~\cite{malinowski2006matching}, who built a bilateral person-job recommendation system using expectation maximization algorithm.
Then, some works casted this problem as a collaborative filtering task~\cite{lu2013recommender,zhang2014research}.

The major limitation of early methods lies in the ignorance of the semantic information of job/resume documents.
Thus, recent research mostly treats person-job fit as a text-matching task.
Several models are designed to extract expressive representations from resume and job description documents via Convolutional Neural Network (CNN)~\cite{zhu2018person},
Recurrent Neural Network (RNN)~\cite{qin2018enhancing} and self-attention mechanisms~\cite{luo2019resumegan,bian2020learning}.
Besides,
rich features of candidates and jobs are also leveraged, such as historical matched documents~\cite{yan2019interview}, multi-level labels~\cite{le2019towards}, multi-behavioral sequences~\cite{fu2021beyond} or other general contextual features~\cite{jiang2020learning}.

Most previous work considers person-job interactions from the \emph{recommendation} scenario as implicit or explicit signals.
As a comparison, we would like to extend this body of research by leveraging search history of users on online recruitment platforms,
which is a valuable resource to understand the job intentions of users. 
We are also aware that there are some general studies that jointly consider search and recommendation~\cite{zamani2018joint,zamani2020learning}.
However, to our knowledge, we are the first to leverage search history for improving person-job fit task.

%% file: sec-def.tex
\section{Problem Definition}\label{sec:def}

Assume there are a set of job positions $\mathcal{J} =\{j_1, j_2, \ldots, j_M\}$
and a set of candidates $\mathcal{U} = \{u_1, u_2, \ldots, u_N\}$,
where $M$ and $N$ represent the total number of job positions and candidates respectively.
Each job  $j$ (job ID) is associated with a text of job description  $t_j$, and each candidate 
$u$ (user ID) is associated with a text of resume $r_u$.
And, an observed match set (recommendations with explicit feedback)
$\mathcal{D} = \{ \langle u, j, y_{u,j} \rangle |\, u \in \mathcal{U}, j \in \mathcal{J}\}$ is given,
where $y_{u,j}$ is the binary label indicating the match result
($1$ for \emph{success} and $0$ for \emph{failure}) for user $u$ and job $j$.

Different from existing settings~\cite{zhu2018person,qin2018enhancing,bian2020learning}, we assume that the search history of each user $u$ is also available. 
On recruitment platforms, a user can issue queries, and then she clicks or applies for some jobs for interviews during the session of this query.
For user $u$, a $L$-length search history is formally denoted as $\mathcal{H}_u = \{\langle q_{1}, j_{1}\rangle, \ldots, \langle q_{L}, j_{L}\rangle\}$, where $q_i$ denotes a query (a short word sequence describing the desired position)
and $j_i$ denotes the job that user $u$ clicks or applies for this job during the query session for $q_i$. 
The tuples in $\mathcal{H}_u$ are sorted in temporal order.
\textcolor{black}{Note that queries in one session might be the same, \ie a user applies for multiple jobs under the same query content.
Besides,} different sessions might correspond to the same query content, \ie a user issued the same query at a different time. 
Based on the observed match set $\mathcal{D}$ and the search history $\mathcal{H}$, our task is to learn a predictive function 
$f(u,j) \in [0, 1]$
to predict the confident score that a user $u \in \mathcal{U}$ will accept job position $j \in \mathcal{J}$ each other.

%% file: sec-model.tex
\begin{figure*}[!htbp]
	\centering
	\includegraphics[width=1\textwidth]{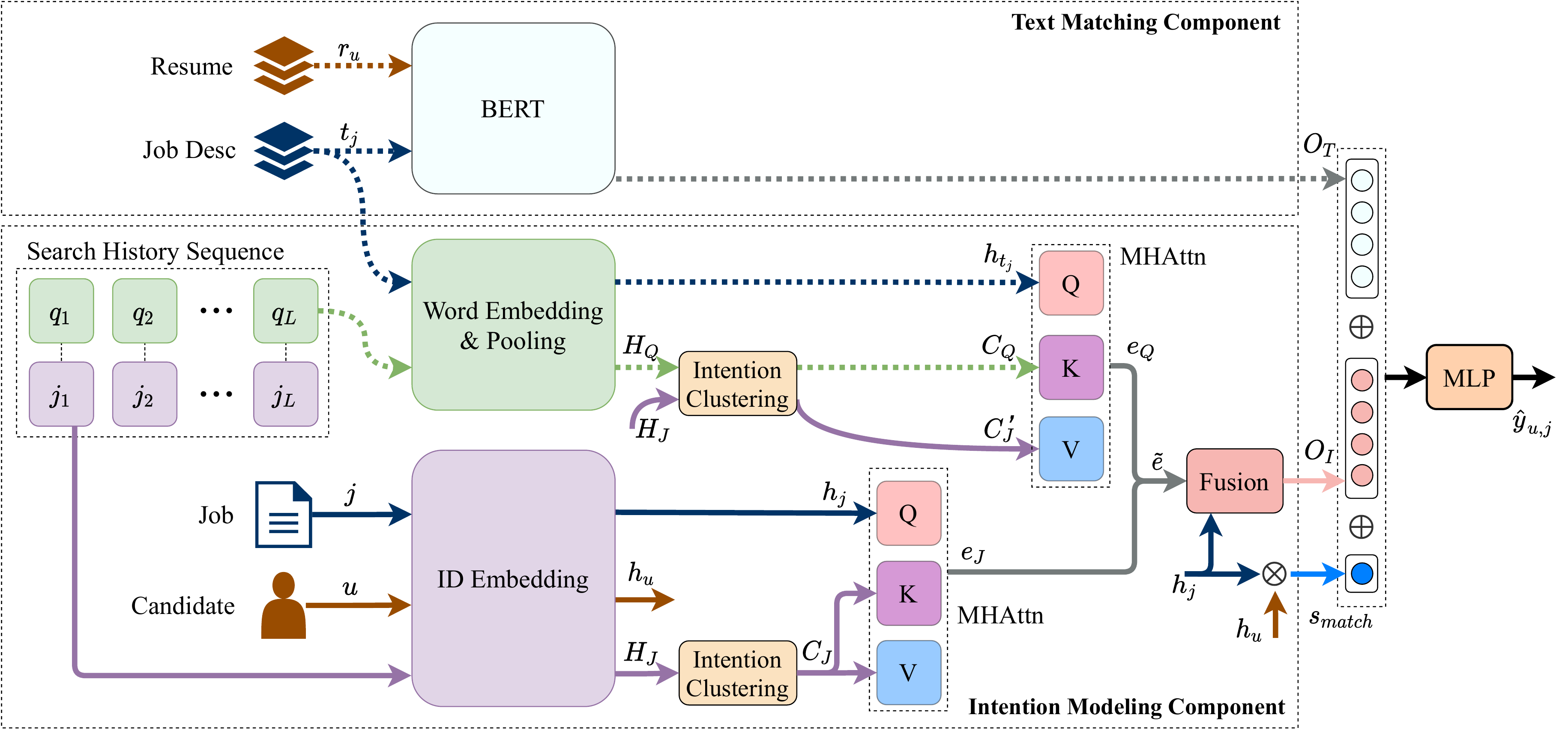}
	\caption{The overall architecture of SHPJF.}
	\label{fig:model}
\end{figure*}

\section{The Proposed Approach}\label{sec:method}

In this section, we present the proposed \textbf{S}earch \textbf{H}istory enhanced \textbf{P}erson-\textbf{J}ob \textbf{F}it model, named as \textbf{SHPJF}. It considers two kinds of data signals to develop the matching model, either text content from resumes/job descriptions or search histories issued by users. On one hand, we design a BERT-based text encoder for capturing the semantic interaction between resumes and job descriptions,
called \emph{text matching component}. On the other hand, we leverage the search history to model the intention of a candidate, called \emph{intention modeling component}.
Based on the Transformer architecture, we design two kinds of intention modeling approaches, either with job ID sequence or with query text sequence. 
Figure~\ref{fig:model} presents the overall architecture of our proposed approach.

\subsection{Text Matching Component}\label{sec:text_matching}
On online recruitment platforms, a major criterion for person-job fit is that a candidate's resume should well match the job description in text contents. It is not easy to learn a good semantic match model for this task, as candidate resumes and job descriptions are likely to be described or presented in different ways~\cite{zhu2018person}. 

Recent years, self-attention mechanisms~(\eg Transformer~\cite{vaswani2017attention})
and its extensions on pre-trained model~(\eg BERT~\cite{devlin2019bert}) have made great progress in various natural language processing tasks.
Therefore, we adopt a standard BERT model to develop the text matching component.
Given a pair of resume $r_u$ and  job description $t_j$,
we firstly concatenate (1) a special symbol \texttt{[CLS]}, (2) resume document,
(3) a special symbol \texttt{[SEP]}, and (4) job description document, in order and derive the input sequence for BERT.
Then the concatenated sequence is fed to the BERT model, so we have
\begin{equation}
    \bm{o}_{T} = \text{BERT}([CLS]; r_u; [SEP]; t_j]),\label{eq:ot}
\end{equation}
where $\bm{o}_{T}$ is the final hidden vector corresponding to the first input token (\texttt{[CLS]}).
The self-attention mechanism in BERT can effectively characterize the semantic interaction between resume and job description.%

Note that another possible architecture is to adopt a two-tower encoder with two separate BERTs~\cite{bian2020learning}. 
However, as shown in previous studies in dense retrieval~\cite{humeau2020poly}, the currently adopted single-tower encoder (also called \emph{cross encoder})
is more effective than the two-tower encoder, since it can capture fine-grained semantic interaction between the texts on two sides.

\subsection{Intention Modeling Component}\label{sec:im}
Above, text-based matching component mainly measures the matching degree between job and candidate based on 
semantic compatibility in content. As introduced in Section~\ref{sec:def}, search history data is also available in our 
setting. Since search history contains the issued queries and the applied (or clicked) jobs, it provides important evidence for modeling user's intention. To learn such intention, we design a two-stream intention aggregation approach, where both job ID sequence and query text sequence are modeled to capture the underlying intention semantics. In particular,  different queries are likely to correspond to the same intention, \eg ``\emph{layer}'' and ``\emph{case source}''. Considering this issue, we further design an intention clustering technique. Finally, the derived representations in two streams are combined as the final intention representation.

\subsubsection{Intention Modeling with Job ID Sequence.}\label{sec:im_with_job_id_seq}
Recall that the search history for user $u$ is given as $\mathcal{H}_u = \{\langle q_{1}, j_{1}\rangle, \langle q_{2}, j_{2}\rangle, \ldots, \langle q_{L}, j_{L}\rangle\}$ \textcolor{black}{in Section~\ref{sec:def}}.  We first consider extracting the intention semantic from the sequence of clicked or applied job IDs, namely $j_1 \rightarrow j_2 \ldots \rightarrow j_L$, where these jobs reflect  the job preference of user $u$.
We first embed the job IDs in the search history: 
\begin{align}
    \bm{h}_{1},\bm{h}_{2},\ldots,\bm{h}_{L} =& \text{IDEmb}(j_1, j_2, \ldots, j_L),\label{eq:idemb}\\
    \bm{H}_{J} =& [\bm{h}_{1};\bm{h}_{2};\ldots;\bm{h}_{L}],
\end{align}
where $\bm{h}_j$ denotes the embedding for job $j$ with the $\text{IDEmb}$  layer. 
Here, we apply a look-up operation to obtain job ID embeddings from the $\text{IDEmb}$  layer. 
\textcolor{black}{Note that IDs of the recommended job and those in the search history are embedded into the same semantic space (by the same look-up table $\text{IDEmb}$ in Eqn.~\eqref{eq:idemb}).}

\paratitle{Intention Clustering.}
As one candidate may have diverse job intentions,
 the embeddings of interested jobs from search history are clustered into several \emph{intentions}.
Formally, we have 
\begin{align}
    \bm{C}_{J} =& \bm{P}_{J} \bm{H}_{J},\label{eq:cls1}\\
    \bm{P}_{J} =& \text{softmax}(\bm{W}_{1}\bm{H}_{J}^\top + \bm{b}_1),\label{eq:cls2}
\end{align}
where $\bm{P}_J$ is a probability assignment matrix that gives the probabilities of a job (from search history) into each cluster, and
$\bm{W}_{1}$ and $\bm{b}_1$ are learnable parameter matrix and bias respectively.
Here, $\bm{C}_{J} \in \mathbb{R}^{k \times d_j}$ is an intention embedding matrix, where each of the $k$ intentions is mapped into a $b$-dimensional vector. 

\paratitle{Job-specific Intention Representation.}
After obtaining the representations of $k$ intentions ($\bm{C}_{J}$) for user $u$, we next 
learn the intention representation of user $u$ given a recommended job position $j$. The basic idea is to 
consider the relevance of job $j$ with each of the learned intention representations. Intuitively, 
a job tends to be accepted by a user if it highly matches with some specific intention of the user. 
We adopt the self-attention architecture~\cite{vaswani2017attention} to characterize this idea by attending job $j$ to each intention embedding. 
To be specific, we adopt multi-head attention operation~\cite{vaswani2017attention}:
\begin{align}
    \text{MHAttn}(\bm{Q}, \bm{K}, \bm{V}) = & [head_1,\dots,head_h]\bm{W}^O,\\
    head_i =& \text{softmax}\left(\bm{Q}\bm{W}_i^Q(\bm{K}\bm{W}_i^K)^\top / \sqrt{D}\right)\bm{V}\bm{W}_i^V,
\end{align}
where $\bm{W}_i^Q,\bm{W}_i^K,\bm{W}_i^V$ and $\bm{W}^O$ are parameter matrices,
and $\frac{1}{\sqrt{D}}$ is the scaling factor.
With the above attention operation,
we  specially designed  \textit{queries} (job embedding), \textit{keys} (intention embedding) and \textit{values} (intention embedding) as:
\begin{align}
    \bm{e}_{J} = \text{MHAttn}(\bm{h}_{j}, \bm{C}_{J}, \bm{C}_{J}),\label{eq:e_id}
\end{align}
where $\bm{h}_{j}$ is the embedding for job $j$ obtained from the $\text{IDEmb}$  layer, and $\bm{e}_{J} \in \mathbb{R}^{d_j}$ denotes the learned intention representation based on job ID sequence. 
\textcolor{black}{By mapping job $j$ (as a \textit{query}) against a sequence of intentions (as \textit{keys}),
each intention is assigned with a relevance score. Those highly relevant intentions (as \textit{values}) will receive a larger attention weight.}
Since the recommended job $j$ has attended to each intention, $\bm{e}_{J}$ encodes important evidence for measuring the match degree between job $j$ and user $u$.

\subsubsection{Intention Modeling with Query Text Sequence.}\label{sec:im_q}
To learn user intention, another kind of important data signal from search history is the \emph{query text}, \ie
$q_1 \rightarrow q_2 \ldots \rightarrow q_L$. Each query is a short sequence of words, reflecting the user intention in job seeking. We follow the similar approach as modeling job ID sequence to model query text sequence. 
Firstly, we apply the look-up operation and the average pooling to represent each query in low-dimensional space: 
\begin{align}
    \widetilde{\bm{h}}_{1},\widetilde{\bm{h}}_{2},\ldots,\widetilde{\bm{h}}_{L} =& \text{Pooling}\big(\text{WordEmb}(q_1, q_2, \ldots, q_L)\big),\label{eq:wordemb}\\
    \bm{H}_{Q} =& [\widetilde{\bm{h}}_{1},\widetilde{\bm{h}}_{2},\ldots,\widetilde{\bm{h}}_{L}],
\end{align}
where $\text{WordEmb}(\cdot)$ is a learnable embedding layer, and 
$\text{Pooling}(\cdot)$ is an average pooling layer that aggregates several vectors into a single representation.

\paratitle{Intention Clustering.} By considering both query text and job ID, we can derive a more comprehensive intention learning approach by following Eqn.~\eqref{eq:cls1} and~\eqref{eq:cls2}:
\begin{align}
    \bm{C}_{Q} =& \bm{P}_Q\bm{H}_{Q},\label{eq:cq}\\
    \bm{C}_{J}' =& \bm{P}_Q\bm{H}_{J},\label{eq:cj2}\\
    \bm{P}_Q =& \text{softmax}(\bm{W}_{2}[\bm{H}_{Q};\bm{H}_{J}]^\top + \bm{b}_2),\label{eq:p2}
\end{align}
where $\bm{C}_{Q} \in \mathbb{R}^{k \times d_j}$ and $\bm{C}_{J} \in \mathbb{R}^{k \times d_j}$
denote the learned intention representations ($k$ intentions) based on query text and job ID, respectively,  
$ \bm{P}_Q$ is a probability assignment matrix of the probabilities of jobs in the learned intentions (clusters),  
and $\bm{W}_{2}$ and $\bm{b}_2$ are learnable parameter matrix and bias respectively.

\paratitle{Job-specific Intention Representation.}
After obtaining the representations of $k$ intentions ($\bm{C}_{J}'$ and $\bm{C}_{Q}$) for user $u$, we next 
learn the intention representation of user $u$ given a recommended job position $j$. 
Following Eqn.~\eqref{eq:e_id}, we still adopt the multi-head attention to learn the intention representation:
\begin{align}
    \bm{e}_{Q} = \text{MHAttn}(\bm{h}_{t_{j}}, \bm{C}_{Q}, \bm{C}_{J}),\label{eq:e_query}
\end{align}
where the representation of job description for $j$ (denoted by $t_{j}$) to be matched\textcolor{black}{, denoted as $\bm{h}_{t_j}$,} is used as \textit{query},
the clustered query representations are used as \textit{key},
and the clustered ID representations are used as \textit{value}.
\textcolor{black}{
    Here the relevance of job $j$ to intentions is defined by the similarity between job description representation $\bm{h}_{t_j}$ and the clustered query text representations $\bm{C}_Q$.
    Intentions associated with highly relevant queries are assigned with high attention weights.
}
The derived $\bm{e}_{Q}$ can represent the learned intention through query text. 
A key point is that we still adopt $\bm{C}_{J}$ as \emph{values}, so that $\bm{e}_{Q}$  and $\bm{e}_{J}$ 
can be subsequently fused. 

\subsubsection{Intention Representation Fusion.}
To combine the above two kinds intention-based representations, 
we apply a weighted linear combination:
\begin{equation}
    \tilde{\bm{e}} = \lambda \bm{e}_{J} + (1-\lambda)\bm{e}_{Q},\label{eq:lambda}
\end{equation}
where $ \bm{e}_{J}$~(Eqn.~\eqref{eq:e_id}) and $\bm{e}_{Q}$~(Eqn.~\eqref{eq:e_query}) are the learned intention representations, and $\lambda$ is a tuning coefficient. 
If a recommended job well matches the job seeking intention of a user, it indicates that such a person-job pair should be more likely to be successful. Following \cite{qin2018enhancing}, we measure the match degree by fusion:
\begin{equation}
    \bm{o}_I = \text{MLP}\big([\tilde{\bm{e}};\bm{h}_{j};\tilde{\bm{e}}-\bm{h}_{j};\tilde{\bm{e}} \circ \bm{h}_{j}]\big),\label{eq:oi}
\end{equation}
where $\bm{o}_I$ encodes the match information about this person-job pair based on intention, 
$\bm{h}_{j}$ is the embedding for job $j$, $\tilde{\bm{e}}$ is the combined intention representation (Eqn.~\eqref{eq:lambda}), 
\text{MLP} is multilayer perceptron stacked with fully connected layers, and ``$\circ$'' denotes the hadamard product operation.

\subsection{Prediction and Optimization}
With the text matching and intention modeling component, 
we finally integrate the two match representations to predict the confident score of a person-job pair:
\begin{equation}
    \hat{y}_{u,j} = \sigma(\text{MLP}\big([\bm{o}_T;\bm{o}_I; s_{match} ]\big)),\label{eq:y}
\end{equation}
where $\hat{y}_{u,j} \in [0, 1]$ indicates the matching degree between  candidate $u$ and job $j$, $\bm{o}_T$ and $\bm{o}_I$ are the learned match representations in Eqn.~\eqref{eq:ot} and \eqref{eq:oi}, respectively. In addition to two match representations, we also incorporate a simple match score based on the user embedding and job embedding, \ie $s_{match} = \bm{h}_j^{\top} \cdot \bm{h}_u$, where $\bm{h}_u$ and $\bm{h}_j$ are obtained from the $\text{IDEmb}(\cdot)$ layer. 

We adopt binary cross entropy loss to optimize our model,
\begin{equation}
    \mathcal{L} = \sum_{\langle u, j, y_{u,j}\rangle \in \mathcal{D}} -\left[y_{u,j}\cdot \log{\hat{y}_{u,j}} + (1 - y_{u,j})\cdot\log(1- \hat{y}_{u,j})\right],\label{eq:loss}
\end{equation}
where we iterate the training dataset and compute the accumulate loss.

\paratitle{Learning.} In our model, various kinds of embeddings ($\text{IDEmb}$ layer in Eqn.~\eqref{eq:idemb}, $\text{WordEmb}$ layer in Eqn.~\eqref{eq:wordemb})
and involved component parameters are the model parameters.
Note that each $\text{MHAttn}$ (Eqn.~\eqref{eq:e_id} and Eqn.~\eqref{eq:e_query})
and intention clustering layers (Eqn.~\eqref{eq:cls2} and Eqn.~\eqref{eq:p2}) are with different parameters.
In order to avoid overfitting, we adopt the dropout strategy with a rate of $0.2$.
More implementation details can be found in \textcolor{black}{Section}~\ref{sec:imp}.

\paratitle{Time Complexity}. For online service, it is more important to analyze online time complexity for a given impression list of $q$ job positions.
Our text matching component requires a time of $\mathcal{O}(l m^2 d_w + l m d_w^2)$, where $d_w$ is the \textcolor{black}{token} embedding dimension, $m$ is the truncated length of input tokens and $l$ is the number of BERT layers.
While the cross-encoder architecture can be efficiently accelerated with an approximately learned dual-encoder (\eg distillation~\cite{lan2020albert}).
We can also accelerate our text matching component with the two-tower architecture described in \textcolor{black}{Section}~\ref{sec:text_matching}.
In this way, the representations of resumes and job descriptions can be calculated offline, and we only need to perform an inner product operation online.

As for our intention modeling component,
we adopt a lazy-update technique to pre-calculate each user's intention representations $\bm{C}_{J}$ (Eqn.~\eqref{eq:cls1}), $\bm{C}_{J}'$ (Eqn.~\eqref{eq:cj2}) and $\bm{C}_Q$ (Eqn.~\eqref{eq:cq}). 
We update user's search history and update $\bm{C}_{J}$, $\bm{C}_{J}'$ and $\bm{C}_Q$ offline every several hours.
Thus, the complexity mainly depends on the calculation of multi-head attention mechanism (Eqn.~\eqref{eq:e_id} and Eqn.~\eqref{eq:e_query}).
Suppose $k$ is the number of clusters and $d_j$ is dimension of clustered intention representations.
The complexity of one single pass of intention modeling component is $\mathcal{O}(k d_j^2)$.
\textcolor{black}{Suppose MLP in Eqn.~\eqref{eq:oi} and Eqn.~\eqref{eq:y} are both $s$ layers, the complexity of intention representation fusion (Eqn.~\eqref{eq:oi}) is $\mathcal{O}(s\cdot d_j^2)$.}
\textcolor{black}{For one pair of candidate and job, the complexity of prediction (Eqn.~\eqref{eq:y}) is $\mathcal{O}(s\cdot (d_w+d_j)^2)$.
The overall complexity of online serving for a session of $n$ jobs is $\mathcal{O}\left(nl m^2 d_w+ nlm d_w^2 + nk d_j^2+ns (d_w+d_j)^2\right)$.}
Such a time complexity can be further reduced with parallel efficiency optimization.

%% file: sec-exp.tex
\section{Experiments}

In this section, we conduct extensive experiments to verify the effectiveness of our model.
In what follows, we first set up the experiments, and then present and analyze the evaluation results.

\subsection{Experimental Setup}

\begin{table}[!t]
    \caption{
        Statistics of the datasets.
        $\overline{L}$ denotes the average length of search history per candidate.
        $\overline{|q|}$ denotes the average number of words per query.
    }\label{tab:dataset}
    \centering
    \begin{tabular}{cccccc}
        \toprule
        \#candidates & \#jobs & \#positive & \#negative & $\overline{L}$ & $\overline{|q|}$ \\
        \midrule
        $53,566$ & $307,738$ & $257,922$ & $2,109,876$ & $16.55$ & $1.50$ \\
        \bottomrule
    \end{tabular}
\end{table}

\paratitle{Datasets.}\label{sec:dataset}
We evaluate our model on a real-world dataset
provided by BOSS Zhipin\footnote{\url{https://www.zhipin.com}}, a popular online recruiting platform.
The records of our dataset are collected from the real online logs
between November 3 and 12 \textcolor{black}{in 2020}.
We anonymize all the records by removing identity information
to protect the privacy of candidates.
We release our code in \url{https://github.com/RUCAIBox/SHPJF}.

There are three kinds of user behavior in our dataset,
called \textit{Accept}, \textit{Apply}, and \textit{Exposure}.
\textit{Accept} means that a candidate and a company reach an agreement on an offline interview.
\textit{Apply} means that a candidate applies for a job.
Generally, it means that the candidate shows a clear intention to the applied job~\cite{le2019towards}.
\textit{Exposure} means that a job position has been exposed to the candidate
but the candidate may not perform further behavior.
To construct the evaluation dataset, all the job-user pairs with \textit{Accept} are considered to be \emph{positive instances}, while those pairs with \textit{Exposure} but without further behavior are considered to be \emph{negative instances}.
Since the number of \textit{Exposure} is huge so that for each positive instance we pair it with several negative instances from the same exposure of a recommendation list.  The ratio of positive and negative instances is approximately $1:8$. Note that all the instances are from the \emph{recommendation} channel, we remove the ones which also appear in the \emph{search} channel for the same user to avoid data leakage.

We sort the records of the selected instances by timestamp.
Records of the last two days are used as validation set and test set respectively,
while the others are used for training. 
For a user, we also obtain her search history (queries with clicked jobs). The number of browsing and clicking behaviors is very large and noisy. Therefore, we only keep jobs with the status of \emph{Apply}.   
We truncate the search history before the timestamp of validation set. 
The statistics of the processed data are summarized in Table~\ref{tab:dataset}.

\paratitle{Baselines.}
We compare our model with the following representative methods:\\
$-$ \textbf{PJFNN}~\cite{zhu2018person} is a convolutional neural network (CNN) based method,
        resumes and job descriptions are encoded independently by hierarchical CNNs.\\
$-$ \textbf{BPJFNN}~\cite{qin2018enhancing} leverages bidirectional LSTM to derive the resume and job description representations.\\
$-$ \textbf{APJFNN}~\cite{qin2018enhancing} proposes to use hierarchical RNNs and co-attention techniques to process the job positions and resumes.\\
$-$ \textbf{BERT}~\cite{devlin2019bert} is a broadly used pre-trained model for learning text representations.
        Here, we adopt the released pre-trained BERT-Base-Uncased model.
        Then it was fine-tuned on our dataset as a sentence pair classification task.\\
$-$ \textbf{MV-CoN}~\cite{bian2020learning} is a BERT-based multi-view co-teaching network that is able to learn from sparse,
        noisy interaction data for job-resume matching.

We finally report metrics on the test set with models that gain the highest performance on the validation set.

\paratitle{Implementation Details.}\label{sec:imp}
Text-matching module is initialized via the BERT-Base-Uncased\footnote{https://github.com/huggingface/transformers}.
The dimensions of word embeddings and ID embeddings are $128$ and $16$ respectively.
The dimension of the hidden state is $64$.
Hyperparameters of baselines are tuned in the recommended range from the original papers.
We select combination coefficient $\lambda$ as $0.6$, number of clusters $k$ as $4$ and number of heads $h$ as $1$.
The Adam optimizer is used to learn our model,
and the learning rate is tuned in $\{0.01, 0.001, 0.00001\}$.
The dropout ratio is tuned in $\{0, 0.1, 0.2, 0.3, 0.5\}$.
We adopt early stopping with patience of $5$ epochs.

\paratitle{Evaluation Metrics.}\label{sec:metric}
Given a candidate,
we tend to rank the more appropriate job positions higher in the recommended list.
Thus, we adopt Grouped AUC (GAUC), Recall@K (R@$\{1, 5\}$) and MRR to evaluate our models.
As the traditional AUC metric doesn’t treat different users differently,
we use GAUC~\cite{zhu2017optimized,zhou2018deep}, which averages AUC scores over users.

\begin{table}[t]
    \centering
    \caption{Performance comparisons of different methods.
    The improvement of our model over the best baseline is significant at the level of 0.01 with paired $t$-test.
    }\label{tab:result}
    \begin{tabular}{c|| c c c c}
        \hline
        Method & GAUC & R@1 & R@5 & MRR \\
        \hline
        \hline
        PJFNN & $0.5313$ & $0.1412$ & $0.5192$ & $0.4025$ \\
        BPJFNN & $0.5343$ & $0.1391$ & $0.5217$ & $0.4008$ \\
        APJFNN & $0.5323$ & $0.1403$ & $0.5185$ & $0.4000$ \\
        BERT  & $0.5449$ & $0.1515$ & $0.5297$ & $0.4129$ \\
        MV-CoN & $0.5463$ & $0.1554$ & $0.5307$ & $0.4165$ \\
        \hline
        SHPJF (ours) & \bm{$0.5785$} & \bm{$0.1630$} & \bm{$0.5516$} & \bm{$0.4297$} \\
        \hline
    \end{tabular}
\end{table}

\subsection{The Overall Comparison}

Table~\ref{tab:result} presents the performance comparison between our model and the baselines
on person-job fit.
Overall, the three methods PJFNN, BPJFNN and APJFNN tend to have similar performance on our dataset.
Furthermore, the two BERT-based methods (\ie BERT and MV-CoN) seem to perform better leveraging the excellent modeling capacities of pre-trained models. Different from previous studies on person-job fit, our negative instances are more strong, \ie they are from the same impression list with the positive instance. So, the text match model should be capable of identifying fine-grained variations in semantic representations. This may explain why BERT-based methods are better than traditional neural network-based methods. Another observation is that MV-CoN is slightly better than BERT by further considering multi-view learning. 

As a comparison, our model achieves the best performance on all the metrics. 
In specific, SHPJF can improve the best baseline's GAUC result
by 3.22\% and 5.89\% absolutely and relatively, respectively.
These results demonstrate the effectiveness of our approach.
In particular, we incorporate a special intention component that learns user intention representations from search history. Such a component is able to enhance the base text matching component (a cross encoder based on BERT), which is the key to performance improvement.

\begin{figure}[t]
    \centering
    \begin{subfigure}[t]{0.48\textwidth}
        \centering
        \includegraphics[width=0.58\linewidth]{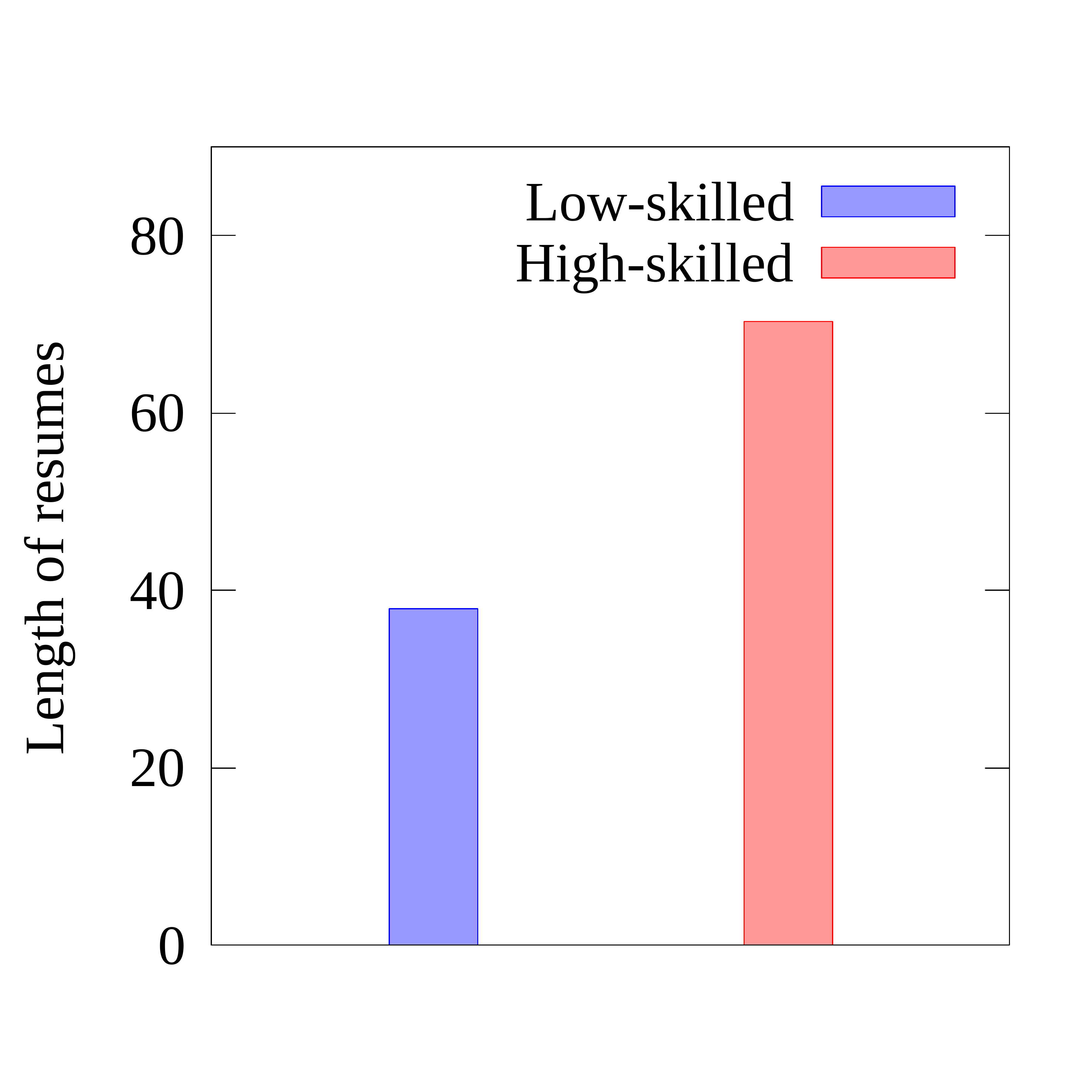}
        \caption{Average resume length.}
        \label{fig:low-skilled:len}
    \end{subfigure}
    \begin{subfigure}[t]{0.48\textwidth}
        \centering
        \includegraphics[width=0.58\linewidth]{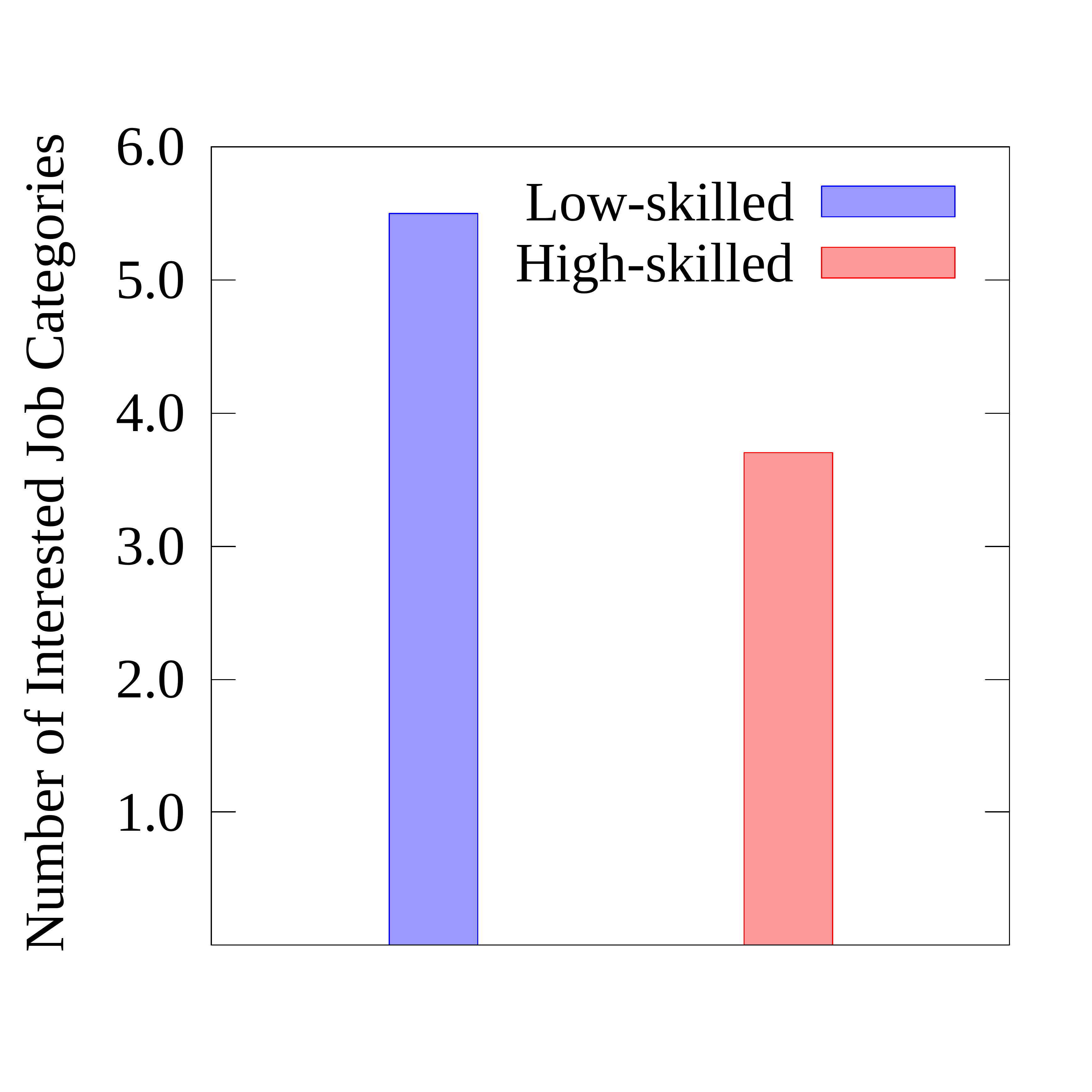}
        \caption{Average \#interested categories.}
        \label{fig:low-skilled:l3}
    \end{subfigure}
    \caption{
        Comparison of low-skilled and high-skilled candidates.
        Each job position belongs to one of the $950+$ pre-defined job categories.
    }
    \label{fig:low-skilled-stat}
\end{figure}

\begin{table}[t]
    \small
    \centering
    \caption{
        Performance comparisons of different methods on low-skilled and high-skilled candidates.
        APJFNN is taken as the base model to compute the RelaImpr.
    }\label{tab:result-improve}
    \begin{tabular}{c||c r||c r}
        \hline
        Groups & \multicolumn{2}{c||}{Low-skilled Candidates} & \multicolumn{2}{c}{High-skilled Candidates} \\
        \hline
        Method & GAUC & RelaImpr & GAUC & RelaImpr \\
        \hline
        \hline
        PJFNN & $0.5295$ & $-4.53\%$ & $0.5318$ & $-2.75\%$ \\
        BPJFNN & $0.5399$ & $+29.13\%$ & $0.5326$ & $-0.31\%$ \\
        APJFNN & $0.5309$ & $0.00\%$ & $0.5327$ & $0.00\%$ \\
        BERT & $0.5381$ & $+23.30\%$ & $0.5470$ & $+43.73\%$ \\
        MV-CoN & $0.5396$ & $+28.16\%$ & $0.5484$ & $+48.01\%$ \\
        \hline
        SHPJF & \bm{$0.5689$} & \bm{$+122.98\%$} & \bm{$0.5814$} & \bm{$+148.93\%$} \\
        \hline
    \end{tabular}
\end{table}

\subsection{Evaluation in Different Skill Groups}

According to the employability in the labor market,
candidates can be divided into \emph{low-skilled} and \emph{high-skilled} candidates.
Furthermore, candidates who applied for jobs
with less training participation and high task flexibility can be classified as low-skilled candidates~\cite{de2004industry,sanders2004training}.
It is usually more difficult for low-skilled candidates to find suitable job positions. 
Therefore, we would like to examine the performance improvement \emph{w.r.t.} different groups.

In our recruitment platform, domain experts manually annotate low-skilled candidates in order to provide specific requirement strategies, so that each user in our dataset will be associated with a label indicating that whether she/he is a low-skilled candidate.
Figure~\ref{fig:low-skilled-stat} shows the average resume lengths and the average number of interested job categories in the two groups. As we can see, low-skilled candidates have a shorter resume in text and a more diverse of interested job categories. These characteristics make it more difficult to apply text-based matching algorithms in finding suitable job positions for low-skilled candidates. As a comparison, our method incorporates search history to enhance the matching model, which is expected to yield larger improvement on low-skilled candidates. 

To examine this, we follow~\cite{zhou2018deep} and introduce \emph{RelaImpr} as a metric of the relative improvements over the base model.
Here, we adopt APJFNN as the base model and check how the other methods improve over it in different groups. 
As shown in Table~\ref{tab:result-improve}, it is more difficult to recommend suitable jobs for  low-skilled candidates (lower performance). The improvement of BERT and MV-CoN is actually very small, which means that BERT-based approaches mainly improve the performance of high-skilled candidates. As a comparison,  our method yields substantial improvement in low-skilled candidates, which further indicates the necessity of leveraging search history data.

\subsection{Ablation Study}\label{sec:ablation}
The major technical contribution of our approach lies in the intention modeling component. 
It involves several parts and we now analyze how each part contributes to the final performance.

\begin{table}[!t]
    \small
    \centering
    \caption{Ablation study of the variants for our model.
    }\label{tab:ablation}
    \begin{tabular}{l|cccc}
        \hline
        Variants & GAUC & R@1 & R@5 & MRR \\
        \hline
        \hline
        BERT & $0.5449$ & $0.1515$ & $0.5297$ & $0.4129$ \\
        $\operatorname{BERT}_{\operatorname{GRU}}$ & $0.5557$ & $0.1546$ & $0.5334$ & $0.4196$ \\
        $\operatorname{BERT}_{\operatorname{query}}$ & $0.5572$ & $0.1558$ & $0.5342$ & $0.4201$ \\
        SHPJF w/o Q & $0.5697$ & $0.1599$ & $0.5456$ & $0.4270$ \\
        SHPJF w/o J & $0.5715$ & \bm{$0.1634$} & $0.5456$ & $0.4286$ \\
        SHPJF w/o C & $0.5738$ & $0.1581$ & $0.5443$ & $0.4237$ \\
        SHPJF & \bm{$0.5785$} & $0.1630$ & \bm{$0.5516$} & \bm{$0.4297$} \\
        \hline
    \end{tabular}
    \hspace{-0.2cm}
\end{table}

We consider the following four variants of our approach for comparison:
\textbf{(A)} \underline{BERT} is a BERT-based text matching model (same as the text matching component in Section~\ref{sec:text_matching});
\textcolor{black}{
\textbf{(B)} \underline{$\text{BERT}_{\text{GRU}}$} replaces the intention modeling component with GRU4Rec~\cite{hidasi2016session} to encode the job ID sequence;
\textbf{(C)} \underline{$\text{BERT}_{\text{query}}$} replaces the intention modeling component (Section~\ref{sec:im}) with a BERT-based model to encode the concatenated query text sequence;
}
\textbf{(D)} \underline{SHPJF w/o Q} removes the part of intention modeling with query text sequence; 
\textbf{(E)} \underline{SHPJF w/o J} removes the part of intention modeling with job ID sequence;
\textbf{(F)} \underline{SHPJF w/o C} removes the intention clustering (Eqn.~\eqref{eq:cls1}, Eqn.~\eqref{eq:cj2} and \eqref{eq:cq}).

In Table~\ref{tab:ablation}, we can see that the performance order can be summarized as:
BERT $<$ \textcolor{black}{$\text{BERT}_{\text{GRU}}$ $\simeq$ $\text{BERT}_{\text{query}}$ $<$} SHPJF w/o Q $<$ SHPJF w/o J $\simeq$ SHPJF w/o C $<$ SHPJF.
These results indicate that all the parts are useful to improve the final performance.
Besides, GRU4Rec doesn't perform well in our experiments.
We observe that user behavior sequences in person-job fit differ from those in traditional sequential recommendation scenarios~\cite{hidasi2016session}.
Existing research also mainly treats person-job fit as a text-matching task but doesn't directly leverage user behavior sequences.
We can see that the carefully designed Transformer architecture performs better in modeling search logs (\eg $\text{BERT}_{\text{GRU}}$ $<$ SHPJF w/o Q, $\text{BERT}_{\text{query}}$ $<$ SHPJF w/o J).
In particular, intention modeling with query text sequences brings more improvement \textcolor{black}{than only with job ID sequences},
as query texts directly reflect users' intention in job seeking.

\begin{figure}[!t]
    \centering
    \begin{subfigure}[t]{0.24\textwidth}
        \centering
        \includegraphics[width=\linewidth]{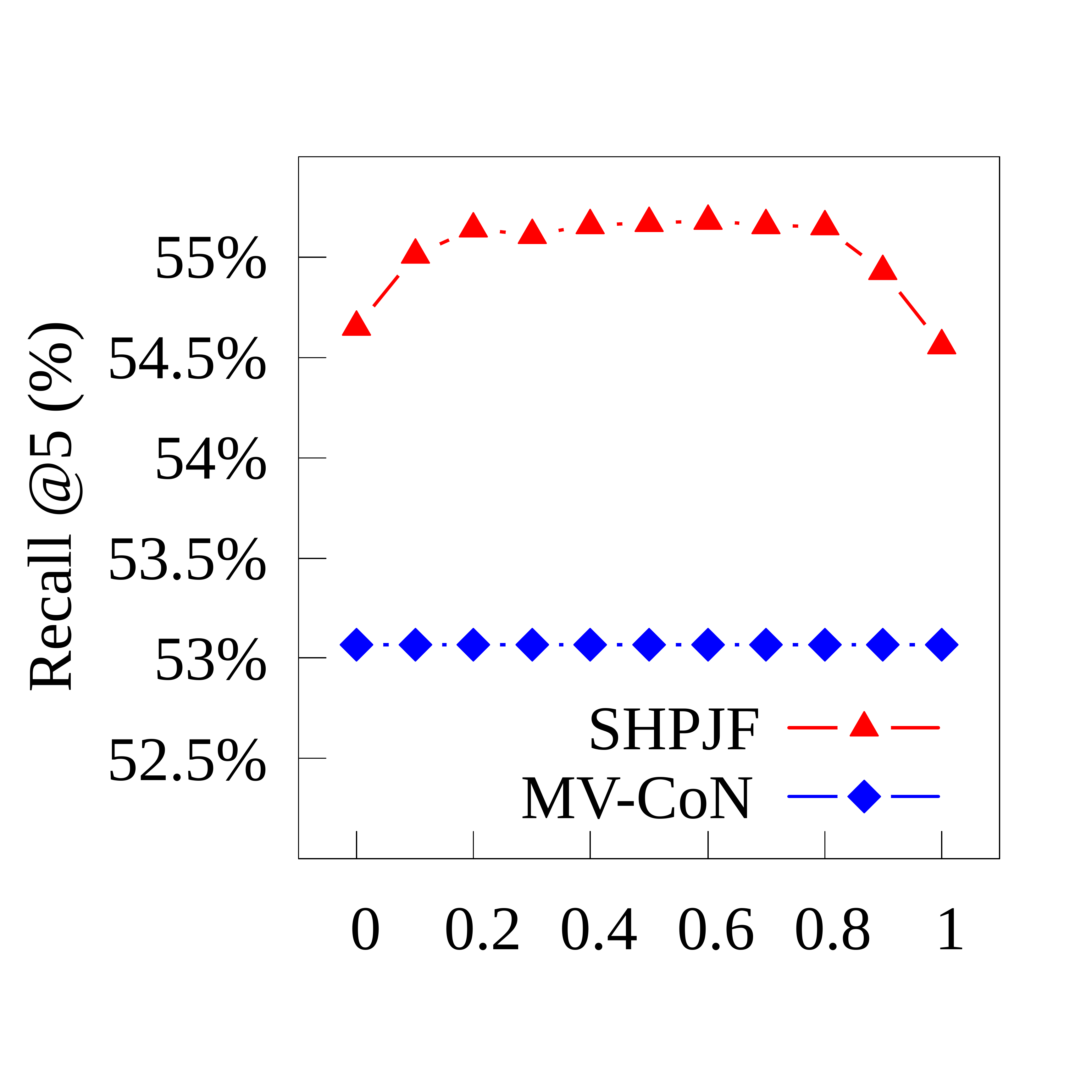}
        \caption{Coefficient $\lambda$.}
        \label{fig:tuning:lambda}
    \end{subfigure}
    \begin{subfigure}[t]{0.24\textwidth}
        \centering
        \includegraphics[width=\linewidth]{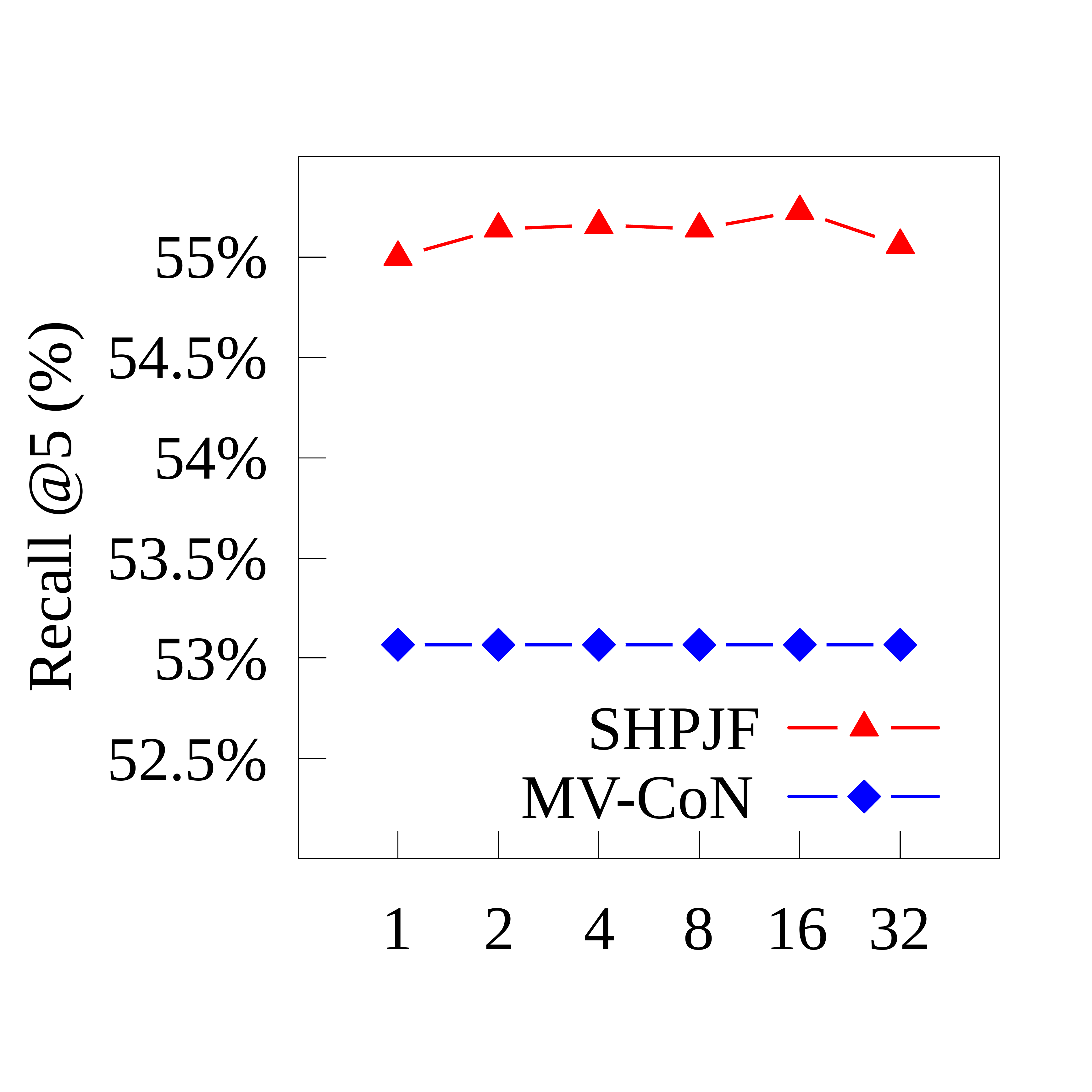}
        \caption{The number of clusters $k$.}
        \label{fig:tuning:C}
    \end{subfigure}
    \begin{subfigure}[t]{0.24\textwidth}
        \centering
        \includegraphics[width=\linewidth]{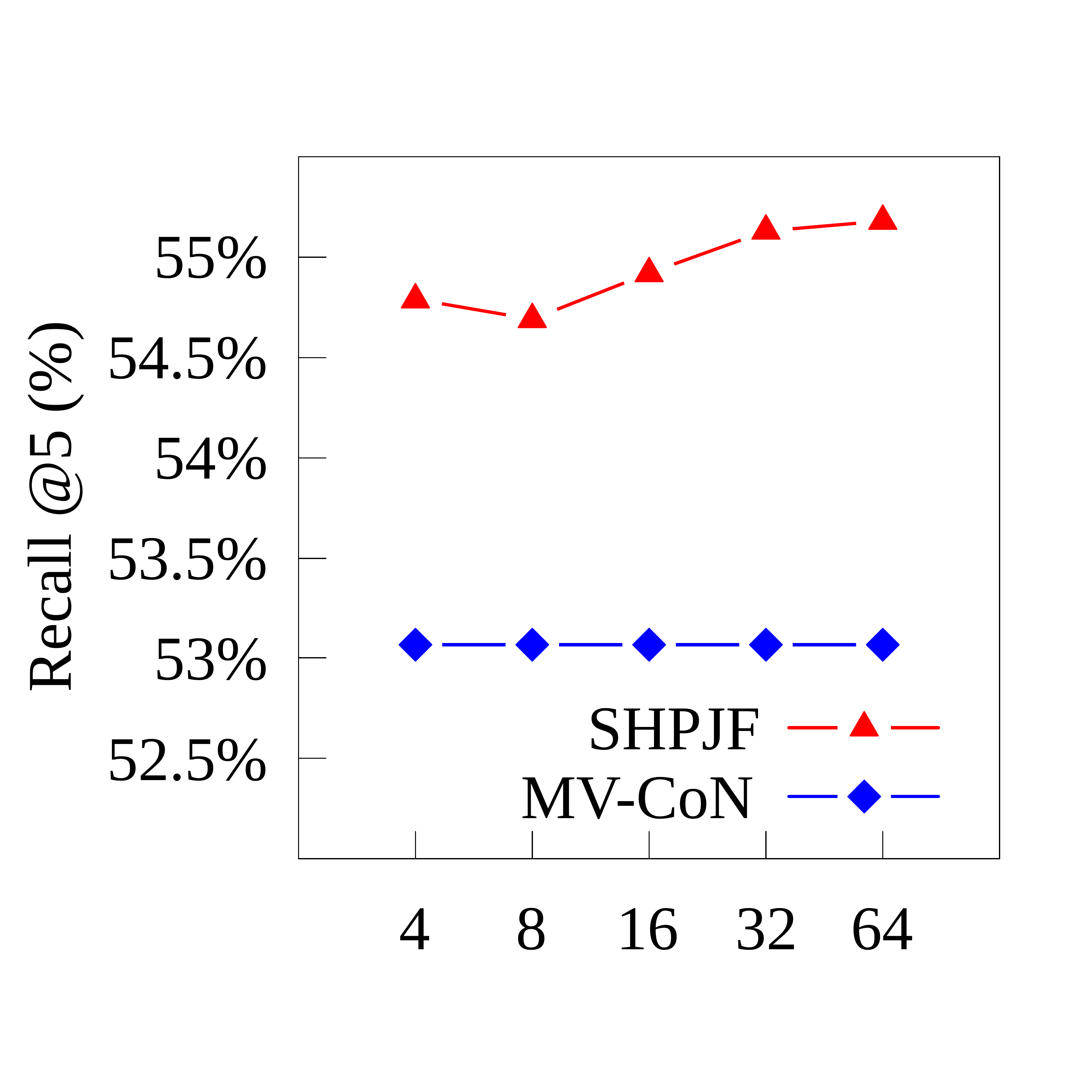}
        \caption{Length of search history sequence.}
        \label{fig:tuning:his_len}
    \end{subfigure}
    \begin{subfigure}[t]{0.24\textwidth}
        \centering
        \includegraphics[width=\linewidth]{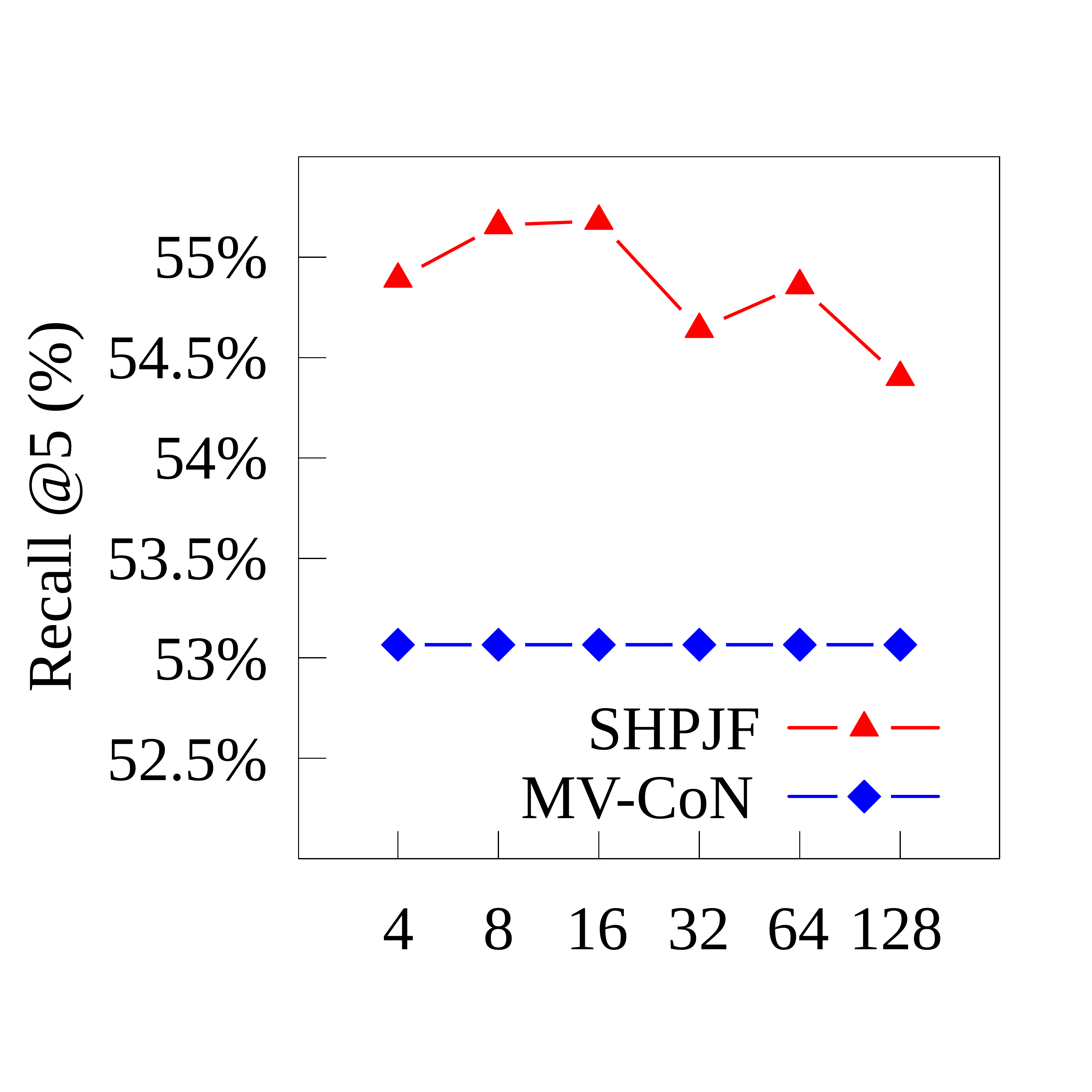}
        \caption{ID embedding dimensionality.}
        \label{fig:tuning:d}
    \end{subfigure}
    \caption{Performance tuning of our model on different hyper-parameters}
    \label{fig:tuning}
\end{figure}

\subsection{Performance Tuning}\label{sec:tuning}

In this part, we examine the robustness of our model and perform a detailed parameter analysis.
For simplicity, we only incorporate the best baseline MV-CoN from Table~\ref{tab:result} as a comparison.

\paratitle{Varying the Combination Coefficient $\lambda$.}
We use a coefficient $\lambda$ to combine the two involved intention representations in Eqn.~\eqref{eq:lambda}.
Here, we vary the combination coefficient $\lambda$ in the range of $0$ and $1$ with a gap of $0.1$.
When $\lambda=1$ and $0$, the model degenerates to SHPJF w/o Q and SHPJF w/o J
described in Section~\ref{sec:ablation} respectively.
As Figure~\ref{fig:tuning:lambda} presented, our model achieves the best performance when $\lambda=0.6$.
With a selection between $0.4$ and $0.8$, the performance is relatively stable.
The results indicate both intention modeling methods are important for our task, which can complement each other in performance.

\paratitle{Varying the Number of Clusters.}
In the intention modeling component, we introduce $k$ to denote the number of clusters. 
The larger $k$ is, the more fine-grained the learned intentions will be. 
Here, we vary the number of clusters in the set $\{1, 2, 4, 8, 16, 32\}$.
As shown in Fig.~\ref{fig:tuning:C}, our model achieves the best performance when $k=4$ and $k=16$.
Overall, the performance is relatively stable with different values for $k$.

\paratitle{Varying the Length of Search History.}
In our work, we leverage search history to enhance person-job fit. It is intuitive that the length of search history will affect the final performance. Here, we vary the (maximum) sequence length of search history in a selection of $\{4, 8, 16, 32, 64\}$.
The tuning results are presented in Fig.~\ref{fig:tuning:his_len}.
From Figure~\ref{fig:tuning:his_len}, we can see that our model gradually improves with the increase of the sequence length. It shows that using more search history will boost the match performance.

\paratitle{Varying the ID Embedding Dimensionality.}
We vary the embedding dimensionality $d_j$ in a selection of $\{4, 8, 16, 32, 64, 128\}$ to examine how the performance changes.
As shown in Figure~\ref{fig:tuning:d}, we find a small embedding dimensionality ($d_j = 8$ or $16$) can lead to a good performance, which can be more efficient in the industrial deployment.

\begin{figure*}[t]
    \centering
    \includegraphics[width=1.0\textwidth]{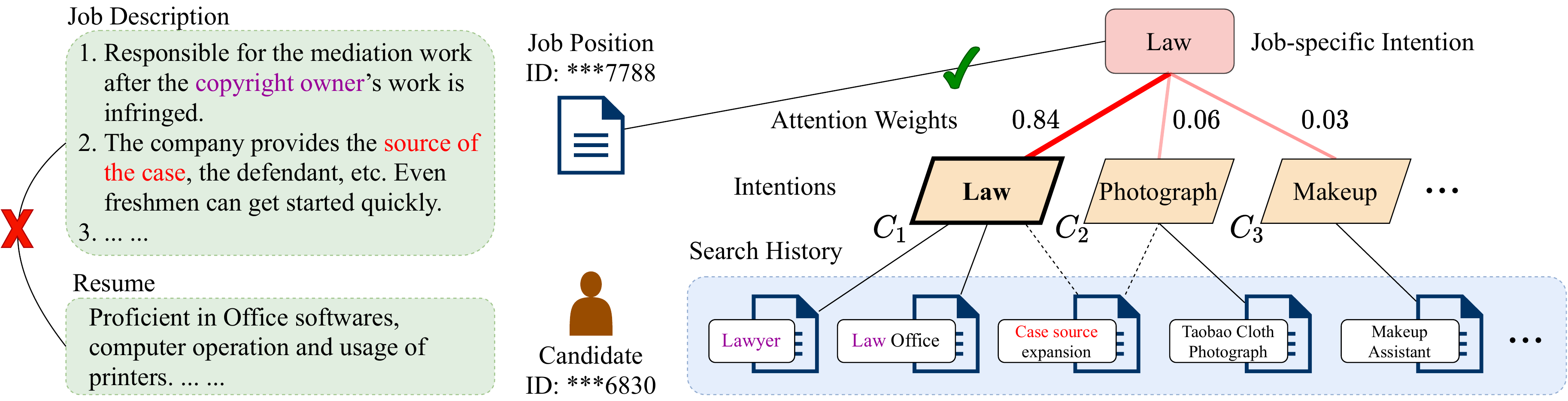}
    \caption{\small Case study: a matched pair between a low-skilled candidate and a job position.}
    \label{fig:case}
\end{figure*}

\subsection{Qualitative Analysis}

In this part, we present a qualitative example, and intuitively illustrate how our model works.
In Figure~\ref{fig:case}, we present a positive case (\ie with the status \emph{Accept} in system) for a user and a job position. 
This interaction record  is randomly sampled from our dataset.
Privacy information has been masked or removed. This user is classified as a \emph{low-skilled candidate} by  a domain expert in our platform. 
As we can see, his resume document is indeed very short, and the job intention is not clearly stated. Given this case, it is difficult to match the candidate with the current position, as the job description has few overlapping words with the resume (semantically different). Therefore, previous text-based matching algorithms~\cite{zhu2018person,qin2018enhancing,bian2020learning} would fail in this matched case.

However, by  checking the candidate's search history,
we find that he has issued queries about several job intentions,
like ``\emph{Makeup Assistant}'', ``\emph{Photograph}'', ``\emph{Lawyer}'', etc.
These intentions cannot be extracted or inferred from his resume. 
While our intention modeling component is effective to derive meaningful clusters about intentions. 
Then, the intention clusters with at least one of the following conditions will be assigned high attentions weights:
\textbf{(1)} contain similar jobs in search history as the job position to be matched.
\textbf{(2)} contain query words highly related or similar to words in the job description, such as ``\emph{source of case - Case source}'' (marked in red) and ``\emph{copyright owner - Law/Lawyer}'' (marked in purple).
Once the correct intention cluster has been identified (with a large attention weight), we can indeed derive a job-specific intention representation (see Section~\ref{sec:im}). In this way, we can correctly match this job-candidate pair even if their documents are semantically different. 

This example shows that leveraging search history is able to improve the performance for person-job fit. It also intuitively explains why our method can yield performance improvement on low-skilled candidates (see Table~\ref{tab:result-improve}).